\documentclass[12pt]{iopart}

\usepackage{iopams}  
\usepackage{setstack}
\usepackage{mymacros_JSTAT}

\usepackage{hyperref}
\newcommand{\doilink}[1]{\url{#1}}
\newcommand{\arXiv}[1]{\url{https://arxiv.org/#1}}

\newcommand{\uzer}{u^{(0)}}

\usepackage{xcolor}

\begin{document}

\title[Brain memory working]{A new mathematical model for brain memory working. Optimal control behavior for Hopfield networks}

%


\author{Franco Cardin$^1$, Alberto Lovison$^2$, Amos Maritan$^3$, Aram Megighian$^4$}
\address{$^1$ Dipartimento di Matematica ``Tullio Levi-Civita'', Università di Padova, Italy }
\address{$^2$ Dipartimento di Matematica, Politecnico di Milano, Italy and Dipartimento di Matematica e Fisica ``Ennio de Giorgi'', Università del Salento, Lecce, Italy }%
\address{$^3$ Dipartimento di Fisica e Astronomia ``Galileo Galilei'', Università di Padova, Italy }
\address{$^4$ Dipartimento di Scienze Biomediche and Padova Neuroscience Center, Università di Padova, Italy}
\eads{\mailto{cardin@math.unipd.it},\mailto{alberto.lovison@unive.it},\mailto{amos.maritan@unipd.it} and \mailto{aram.megighian@unipd.it},}
\vspace{10pt}
%

%


\begin{abstract}
Recent works have highlighted the need for a new dynamical paradigm in the modeling of brain function and evolution. Specifically, these models should incorporate non-constant and asymmetric synaptic weights \(T_{ij}\) in the neuron-neuron interaction matrix, moving beyond the classical Hopfield framework. Krotov and Hopfield proposed a non-constant yet symmetric model, resulting in a vector field that describes gradient-type dynamics, which includes a Lyapunov-like energy function. Firstly, we will outline the general conditions for generating a Hopfield-like vector field of gradient type, recovering the Krotov-Hopfield condition as a particular case. 
Secondly, we address the issue of symmetry, which we abandon for two key physiological reasons: (1) actual neural connections have a distinctly directional character (axons and dendrites), and (2) the gradient structure derived from symmetry forces the dynamics towards stationary points, leading for every pattern to a recognition or to a free association, if the equilibrium is rather far from the input. We propose a novel model that incorporates a set of limited but variable controls \(|\xi_{ij}|\leq K\), which are used to adjust an initially constant interaction matrix, \(T_{ij}=A_{ij}+\xi_{ij}\). Additionally, we introduce a reasonable controlled variational functional for optimization. This allows us to simulate three potential outcomes when a pattern is submitted to the learning system: (1) if the dynamics converges to an existing stationary point without activating controls, the system has \emph{recognized} or has made \emph{a free association} to an incoming pattern; (2) if a new stationary point is reached through control activation, the system has \emph{learned} a new pattern; and (3) if the dynamics \emph{wanders} without reaching any stationary point, the system is unable to recognize or learn the submitted pattern. An additional feature (4) models the processes of \emph{forgetting and restoring} memory. Numerical simulations on a basic neural network model support the theoretical results proposed. 


%
%
\end{abstract}


\section{Introduction}\label{sec1}

The renowned Hopfield model\footnote{Important precursors include W.S. McCulloch and W. Pitts (1943) \cite{McCP}, D.O. Hebb (1949) \cite{He}, and E.R. Caianiello (1961) \cite{C}.} for neural networks \cite{H} consists of $N$ neurons, each with an initial electric status $V_i^{(0)} = 0,1$ ($1$: firing, $0$: not firing), interacting through synaptic weights $T_{ij}$, representing the strength of connection. At each iteration, the potential of neuron $i$, 
\[
u_i^{(k+1)} = \sum_j T_{ij} V_j^{(k)}, 
\]
is calculated by combining the states of other neurons. This is normalized by a threshold activation function $g_0$: 
\[
V_i^{(k+1)} = g_0\left(\sum_j T_{ij} V_j^{(k)}\right), 
\]
where $g_0(x) = 1$ if $x \geq a$ and $g_0(x) = 0$ otherwise. The network stabilizes to a fixed state $V_i^{(\infty)}$ after several iterations.
Nicolas Brunel\footnote{Duke Department of Neurobiology, Durham NC.}, in a 2022 VIMM conference, emphasized the need for generalized Hopfield models where the synaptic weights $T_{ij}$ depend on potentials $V_i$, rather than being constant \cite{B}.
In order to prepare a continuous  version of the above discrete dynamics, 
we first consider an regularized threshold activation function in the form of a sigmoid: $g(x)= g_\epsilon(x) := (1/\pi)\arctan(x/\epsilon)+1/2$,\footnote{See also the resume  \ref{sec:resume} for more details. A repeated index implies summation over its range.} and we set $V=g(u)$. Then we consider two candidate difference schemes: 
%
\begin{equation}\label{pre-ode} \fl
	u_i^{(k+1)}- u_i^{(k)}  = T_{ij}g(u_j^{(k)})- u_i^{(k)},\qquad V_i^{(k+1)}- V_i^{(k)}  = g(T_{ij}V_j^{(k)})- V_i^{(k)}. 
\end{equation}
Although both schemes converge to the same fixed points, we will adopt  (\ref{pre-ode})$_{1}$  in consistency with existing literature. The continuous dynamics is then given by: 
\begin{equation}\label{ode} 
	\dot u_i = T_{ij} g(u_j) - u_i =: X_i(u). 
\end{equation}
In some cases, we will use functions of $V$, like the energy $E(V)$, for algebraic simplicity.

A key feature of Hopfield models is the symmetry of the matrix $T_{ij}$, ensuring the network follows gradient dynamics with a Lyapunov-type energy $E(V)$.
Krotov and Hopfield \cite{K,K-H} extended this idea by introducing a non-constant, Hessian form synaptic matrix:
$ T_{ij} (V) = \nabla_{ij}^{2} \Phi (V)\footnote{Notation for derivatives: for brevity, here and everywhere in what follows, we will write:
\begin{equation*}\fl
\grad_{j}f(x) := \frac{\partial f}{\partial x_{j}}(x), \quad  \grad f(x) :=\tonde{ \grad_{1}f(x),\dots, \grad_{n} f(x)},\quad\text{and }\quad \grad^{2}_{i j}f(x) := \frac{\partial^{2} f}{\partial x_{i}\partial x_{j}}(x), 
\end{equation*}
for every multivariate function $\R^{n}\supseteq U\owns x \mapsto f(x) \in \R$, $f(x) = f(x_{1},\dots,x_{n})$.
}  $
preserving the gradient-like structure and energy properties (see (\ref{eq:grad_nohat}-\ref{eq:grad_hat}) and (\ref{eq:energy_class})).
In Section \ref{sec:prop_uno}, we show that Krotov's hypothesis is a special case of a more general condition (\ref{es}) descending a closure condition on differential forms, leading to gradient dynamics.

The second part of our study (Section \ref{sec:oc}) questions the assumption of symmetry in $T(V)$ for two reasons:
\begin{enumerate}
    \item \textbf{Physiological basis:} Neural connections are directed (from axon to dendrite), so the synaptic matrix $T_{ij}$ should not be symmetric ($T_{ij} \neq T_{ji}$).
    \item \textbf{Questionable gradient structure:} A global Lyapunov function implies that every input pattern is brought towards an equilibrium, which represents either recognition of the pattern, or the attainment of a different pattern, that we will call a ``free association''. This contradicts the real brain behaviour, where many patterns are not recognised nor recorded.
\end{enumerate}
To address these concerns, we assume that $T_{ij} = A_{ij} + \xi_{ij}$, where $A_{ij}$ is a constant matrix and $\xi_{ij}$ are small correctors, and analyze this in an optimal control framework (Section \ref{sec:oc}). This approach unifies four behavioral outcomes:
\begin{enumerate}
    \item Recognition of the initial pattern or free association to a different pattern without control activation.
    \item Recording of a new pattern with control activation.
    \item Failure to achieve equilibrium (non-recognition).
    \item Forgetting and restoring memories.
\end{enumerate}
The role of asymmetry in Hopfield networks has been studied extensively, from early work \cite{Derrida:1987aa, Kanter:1988aa, Parisi:1986aa} to recent developments \cite{Leonetti:2020aa, Rozier:2023aa}. Asymmetry enhances efficiency, supports oscillations, and metastable states like confusion or wandering behavior, which are linked to associations between different memories \cite{Jinwen:1993aa, Zheng:2010aa}. Oscillations arise in non-gradient fields, as noted in \cite{Yan:2013aa}, and are a well-known feature of brain activity, supporting our non-gradient approach.  

This perspective introduces a variational bridge between the old and new synaptic matrices via optimization, incorporating Parisi's suggestion \cite{P1} on bounded synaptic variations. Optimal control theory offers a framework to explore biological neural network behaviors, though further physiological and numerical research is required.



Our task here has been to naturally frame the brain's optimization behavior within a well--known class of problems, i.e., the Infinite Horizon Optimal Control. Detailed existence conditions for such variational processes are thoroughly covered in literature  \cite{BCD,BP,Todorov:2006aa}. 
Explorations analogous to our proposal, involving control theory with a least-action principle, have been applied 
recently in the context of recurrent neural networks \cite{Meulemans:2022aa} and in cortical processes of sensory streams \cite{Senn:2024aa} obtaining competitive computational performances.
In our proposal,  we exhibit a proof of concept of the predicted behavior classes described above (recognition, recording, forgetting and restoring, wandering) by means of numerical simulations on a simple network model in Section \ref{sec:numerics}.

\section{The Krotov extension to the Hopfield model}\label{sec:prop_uno}

%
A popular way of updating dynamically a network is the celebrated proposal by  Hebb  \cite{He}
%
\[
	T_{ij}^{\mathrm{old}}\longrightarrow
	T_{ij}^{\mathrm{new}}=T_{ij}^{\mathrm{old}}+\frac{1}{N}{\widehat{V}}_i\,{\widehat{V}}_j,
\]
where the tensor product $\frac{1}{N}{\widehat{V}}_i\,{\widehat{V}}_j $ introduces in the dynamics a new equilibrium state $\widehat{V}$.  

\begin{figure}[htbp]
\begin{center}
	\includegraphics[width=88mm]{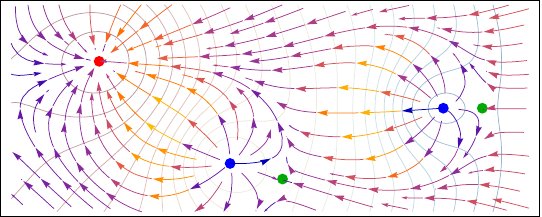}
\caption{Gradient dynamics example: the streamlines are the steepest descent paths of the energy landscape (contours). Blue dots are the maxima (sources/unstable equilibria), red dots are the minima (sinks/stable equilibria) while the green dots are the saddles (unstable equilibria).}
\label{fig:gradientdynam1}
\end{center}
\end{figure}

Recently, Krotov and Hopfield \cite{K, K-H} have proposed an extension of the classical Hopfield model in the case where the synaptic  matrix $T_{ij}$ may depend, although 
still in the symmetric case, on the values of the electric status $V_i$ with the apparently artificial structure
\begin{equation}\label{a}
T_{ij}(V)=\nabla_{ij}^2\Phi(V),   
\end{equation}
for a real valued function $\Phi$.
%
%
In our opinion the interesting underlying idea leading to the proposal
of the special structure (\ref{a}) is the request that the vector field $X_i(u)$, with $T_{ij}=T_{ij}(V)$, is of gradient type, i.e., in the form:
\begin{equation}\label{eq:grad_nohat}
\dot u_i = T_{ij}g(u_j)-u_i=X_i(u)=-\nabla_{i} f(V)|_{V=g(u)}\,,
\end{equation}
or
\begin{equation}\label{eq:grad_hat}
{\widehat X}_i(V)=X_i(u)|_{u=g^{-1}(V)}=-\nabla_{i} f(V)\,,
\end{equation}
for a suitable real valued function $f(V)$ (see below (\ref{eq:krotov-energy})),
which assumes the role of Lyapunov energy function for the asymptotic behavior:
\begin{eqnarray}\label{dotf} \fl
\dot f=\nabla_{i} f(V)|_{V=g(u)} \dot V_i= \nabla_{i} f(V)|_{V=g(u)}   \frac{d}{dx}     g(x)|_{x=u_i}\dot u_i  =  -  \frac{d}{dx}     g(x)|_{x=u_i} X_i  (u) ^2 \leq 0\,.
\end{eqnarray}
The trajectories following the vector field are the steepest descent paths of the energy represented with the underlying contour plot, as shown in Figure in \ref{fig:gradientdynam1}.
%
Let us examine \cite{K, K-H} more closely. We first revisit the energy in the classical Hopfield model (see Section \ref{sec:resume} for a resume), where
\[T_{ij} = A_{ij} \ \ \mathrm{with} \ \  A_{ij}\   \mathrm{constant.}\] 
In this case we have $f(V)=E_A(V)$, defined as:
\begin{equation}\label{eq:energy_class}
	E_A(V)= -\underbrace{\frac{1}{2}\sum_{i,j=1}^N A_{ij}V_iV_j}_{{\cal J}_A(V), \ \mathrm{ see }\ (\ref{444})\ \mathrm{ below.}}+\sum_ {i=1}^N \int_0^{V_i}g^{-1}(x)dx\,.
\end{equation} 
Defining the function $\Phi_A(V)$ as the dominant term of the energy:
\begin{equation}\label{2} \Phi_A(V) := \frac{1}{2} A_{ij} V_i V_j, \end{equation}
we find:
\begin{equation}\label{3} \nabla_{ij}^2 \Phi_A(V) = A_{ij}. \end{equation}
This indicates that $\Phi_A$ corresponds to the Jacobi function ${\cal J}_A$ related to the Lagrangian $\Phi_A$:
\begin{equation}\label{444} 
	{\cal J}_A(V) = \nabla \Phi_A(V) \cdot V - \Phi_A(V) = \frac{1}{2} A_{ij} V_i V_j. 
\end{equation}
The authors of \cite{K, K-H} extend the classical Hopfield model by introducing a more general Lagrangian $\Phi(V)$ and 
extending the above setting (\ref{3}):
\begin{equation}\label{6} \nabla_{ij}^2 \Phi(V) = T_{ij}(V). \end{equation}
This new $\Phi(V)$ can be viewed as a perturbation of (\ref{2}):
\begin{equation}\label{eq:new_lagrangian} 
	\Phi(V) = \frac{1}{2} A_{ij} V_i V_j + \varphi(V), 
\end{equation}
for some scalar function $\varphi(V)$.
The related dynamics now becomes:
\begin{equation}\label{7-1}
\dot{u}_{i} = \widehat{X}_{i}(V)\big|_{V = g(u)} = \left( \nabla_{ij}^2 \Phi(V) V_j - g^{-1}(V_i) \right) \big|_{V = g(u)}. \end{equation}
By analogy with the constant matrix case, we reformulate the Jacobi function (\ref{444}) and the energy function (\ref{eq:energy_class}) based on the new Lagrangian $\Phi(V)$. The Jacobi function is now:
\begin{equation}\label{70} {\cal J}(V) = \nabla \Phi(V) \cdot V - \Phi(V), \end{equation}
and the energy becomes:
\begin{equation}\label{eq:krotov-energy} 
E(V) = -\left( {\cal J}(V) - \sum_i \int_0^{V_i} g^{-1}(x) , dx \right). \end{equation}
We can verify that $f (V)= E(V)$:
\begin{equation} -\nabla_i E(V) = \nabla^2_{ij} \Phi(V) V_j - g^{-1}(V_i) = X_i(u) \big|_{u = g^{-1}(V)}, \end{equation}
and that the Lie derivative $\dot E$ behaves as in (\ref{dotf}) above.



\section{Proposal one: gradient vector field}\label{sec:prop_uno}

We have shown that the proposal (\ref{6}) by Krotov and Hopfield, while initially seeming unnatural, leads to a gradient dynamics and a Lyapunov energy $E(V)$. Here, we explore more general conditions that can yield the gradient structure.

\begin{prop}\label{prop:uno} 
Let the vector field $\widehat{X}(V)=X(u)|_{u=g^{-1}(V)}$ generate the dynamics (\ref{ode}) with a symmetric $T_{ij}(V)=T_{ji}(V)$. The standard condition to be a \emph{gradient vector field} in a simply connected domain, i.e.\ $ \nabla_{i} \widehat{X}_j = \nabla_{j} \widehat{X}_i$, is equivalent to:
\begin{equation}\label{es} \tonde{\grad_{i}T_{kj} - \grad_{j}T_{ki}} V_k = 0. \end{equation} 
\end{prop}
\begin{proof} Expanding $ \nabla_{i} \widehat{X}_j = \nabla_{j} \widehat{X}_i$, we obtain:
$$
       \nabla_{i} (T_{jk}(V)V_k)= \nabla_{j}(T_{ik}(V)V_k),\qquad 
        (\grad_{i}T_{jk})V_k+T_{ji}= (\grad_{j}T_{ik})V_k+T_{ij}.
$$
%
Given the symmetry of $T_{ij}$, we obtain (\ref{es}). 
\end{proof}


\begin{rem}\label{rem:uno} 
If we assume the stronger (than (\ref{es})) condition:
\begin{equation}\label{ess} 
	\grad_{i}T_{kj} = \grad_{j}T_{ki}, 
\end{equation}
for some vector $Z_k$ we have 
$$
	T_{kj}=\grad_{j}Z_{k}=\grad_{k}Z_{j}\,,
$$ 
for the symmetry of $T_{kj}$. 
Consequently, there exists a scalar function $\Phi$ such that $Z_k =\grad_{k}\Phi$ and its Hessian is exactly $T_{ij}$, i.e.:
\begin{equation}\label{66} 
	\grad^{2}_{ij}\Phi(V) = T_{ij}(V), 
\end{equation}
restoring the original hypothesis by Krotov and Hopfield (\ref{6}). \qed 
\end{rem}

\begin{rem} While condition (\ref{ess}) implies (\ref{es}), the converse does not hold. 
Indeed, consider the symmetric matrix\footnote{This example is due to Giuseppe De Marco.} for $N=2$:
$$
T(V)=\left(\begin{array}{cc}0&V_1 V_2\\ V_1 V_2&0\end{array}\right)\,.
$$
Here, (\ref{ess}) does not hold since:
$$
\grad_{2}T_{11}-\grad_{1}T_{12}=-V_2\neq 0,\qquad \grad_{2}T_{21}-\grad_{1}T_{22}=V_1\neq 0,  $$
i.e., $\grad_{i}T_{kj}\neq \grad_{j}T_{ki}.$ However, 
$$
(\grad_{i}T_{kj} - \grad_{j}T_{ki})V_{k} = (\grad_{2}T_{11}-\grad_{1}T_{12})V_1 + (\grad_{2}T_{21}-\grad_{1}T_{22})V_2=0\,,
$$
so (\ref{es}) holds. \qed 
\end{rem}
Under the assumption of (\ref{es}) in Proposition \ref{prop:uno}, we construct the energy  $E(V)$, which is a primitive of the exact differential form $-\widehat X(V)dV$:
\begin{equation}
    E(V) := -W(V) + \sum_{i=1}^N \int_0^{V_i} g^{-1}(a)da, 
\end{equation}
where
\begin{equation}
W(x):=\int_0^1 T_{ij}(\lambda x)\lambda x_i x_j d\lambda.
\end{equation}
Thus, we obtain:
$$
\widehat{X}_i = T_{ij}(V)V_j - g^{-1}(V_i) = -\nabla_i E(V). 
$$
Differentiating $W$ and using the symmetry of $T_{ij}$, we find:
\begin{equation} \eqalign{\fl 
\grad_{k} W(x)=  
\int_0^1 \Big(\grad_{k} T_{ij}(\lambda x) \lambda^2 x_i x_j + T_{kj}(\lambda x) \lambda x_j  +T_{ik}(\lambda x) \lambda x_i  \Big)d\lambda,   \\ 
=\int_0^1 \Big(\grad_{k} T_{ij} (\lambda x) \lambda^2 x_i x_j + 2T_{ik}(\lambda x) \lambda x_i \Big)d\lambda  \\
\overset{\mathrm{because\ of\ (\ref{es})}}{=} 
\int_0^1 \Big(\grad_{j} T_{ik} (\lambda x) \lambda^2 x_i x_j +2T_{ik}(\lambda x) \lambda x_i  \Big)d\lambda   \\
= \int_0^1 \frac{d}{d\lambda}\Big(T_{ik} (\lambda x)\lambda^2  x_i \Big) d\lambda=T_{ik} (x) x_i  
=T_{ki} (x) x_i.}
\end{equation}
Thus, the gradient of $-E$ corresponds to the desired vector field $\widehat X$. As a result, $E(V)$ serves as a true Lyapunov function:
\begin{equation}
\dot{E}(V) = \nabla_{j}E(V)\dot V_j=-\dot u_j\dot V_j=-\underbrace{\frac{d}{dV}g^{-1}(V_j)}_{>0} ({\dot V}_j)^2\leq 0.
\end{equation}
\phantom{.}\hfill \qed



\section{Proposal two: an alternative perspective through Optimal Control}\label{sec:oc}
Within the Hopfield framework, we address the challenges posed by a non-constant and non-symmetric synaptic matrix $T_{ij}$ using a basically new approach.
We propose a constitutive structure for the synaptic matrix defined as:
\begin{equation}\label{constitutive}
    T_{ij}(\xi)=A_{ij}+ \xi_{ij}
\end{equation}
where $A_{ij}$ is a constant (not necessarily symmetric) ``ancestral'' synaptic matrix and $\xi_{ij}$ are bounded adjustments satisfying:
\begin{equation}\label{bound}
    |\xi_{ij}|\leq K. 
\end{equation}
%
The network dynamics is modelled as an optimal control problem 
as follows: 
%
\begin{equation}\label{dynamics} \fl
   \dot u_i(t)= X_i(u(t),\xi(t))=   \sum_{j=1}^N\big( A_{ij}+\xi_{ij}(t)\big)g(u_j(t))- u_i(t), \qquad u_i(0)=u^{(0)}_i,
\end{equation}
where the controls $\xi(t)$ are chosen to minimize a suitable ``economy principle'' discussed below (\ref{eq:varprin}). 

It is well known that the neuron-neuron connections in the brain are not a fully connected network. We therefore  adopt a more realistical model consisting in a 
sparse network where the set of active connections $C=\set{(i,j)| A_{ij}\neq 0}$ is significantly smaller than the total possible connections (i.e., the network is sparse $\# C \ll N\times N$). Recent studies \cite{V} suggest modulating $\xi_{ij}$ between synaptic connections in a set $M\supsetneq C$. This approach naturally accounts for \emph{silent synapses} ($A_{jk}=0$ but possibly $\xi_{jk}\neq 0$), as discussed in the context of brain plasticity \cite{V}.

Given the infrequent activation of silent synapses, we propose a stricter bound on the strength of the new connections:
\begin{equation}\label{bound1}
 \mathrm{for}\ (i,j)\in M\setminus C:
|\xi_{ij}|\leq k\ll K.
\end{equation}
Estimates suggest that $N\approx 10^{11}$ and $10^{13} \,\lesssim \, M \,\lesssim \, 10^{15} \ll N\times N \cong 10^{22}$.

We unify the constraints (\ref{bound}) and (\ref{bound1}) by writing:
\begin{equation}\label{eq:bound2}
    |\xi |\leq (K,k)\,.
\end{equation}
The variables $\xi_{ij}$ are viewed as controls that perturb and modify the ancestral matrix $A_{ij}$, enabling the brain to move the initial pattern $V^{(0)}=g(u^{(0)})$ towards equilibrium (subject to (\ref{eq:bound2})). 

The above mentioned ``principle of economy'' governing the new dynamics is:

\begin{defin}[Infinite Horizon Optimal Control Problem (\cite{BCD,BP})]

For a chosen $\lambda>0$, consider the $e^{-\lambda t}-$\emph{discounted} variational principle:
\begin{equation}\label{eq:varprin} \fl
	\inf_{ |\xi(t)|\leq (K,k)}  J(u^{(0)},\xi) = 
	\inf_{ |\xi(t)|\leq (K,k)} \int_0^{+\infty} 
	\left(|X\big(u(t,u^{(0)}, \xi), \xi (t)\big)|^2+|\xi(t)|^2\right)e^{-\lambda t}\,dt,
\end{equation}
where, for any control $\xi:[0,+\infty[\ni t\mapsto |\xi(t)|\leq (K,k)$, the curve $u(t,u^{(0)}, \xi)$ satisfies the dynamics in (\ref{dynamics}). 

\end{defin}

The integrand function $\ell(u,\xi)$:
\begin{equation}
	\R^N\times \R^{ M}\ni    (u,\xi)\longmapsto \ell(u,\xi):=| X(u, \xi)|^2+| \xi|^2\in{ \R}^+, 
\end{equation}
is referred to as the \emph{Lagrangian function} of the control problem. The discount factor $e^{-\lambda t}$ ensures convergence.

%
%

%

\begin{rem} The variational principle translates the brain's mechanism of searching for or constructing equilibria with minimal deviations from the original synaptic conductivities. We extend the time to the entire interval $[0,+\infty)$ because, typically, achieving equilibrium may require infinite time. However, practically, we can approximate the equilibrium very quickly. For instance, consider the equation $\dot x=-\lambda x$. If the initial condition $x_{0}$ is $\varepsilon-$close to equilibrium $x=0$, then we reach $\varepsilon^2$-closeness to $x=0$ in a time proportional to $\ln ( 1/\varepsilon)/\lambda$. \end{rem}

\begin{rem}[Taxonomy of Possible Network Behaviors]\label{rem:taxonomy} Let $t\mapsto \uzer$ be the curve minimizing the functional (\ref{eq:varprin}). Such resulting optimal controlled dynamics may exhibits various behaviors:
\begin{enumerate}[($i$)] 
\item $u(t,u^{(0)},\xi)$ may converge to a pattern $u_\infty$ 
without the activation of the controls, i.e., with $\xi(t) \equiv 0$. Then $u_\infty$ is an equilibrium also of the original dynamics, indeed  $\lim_{t\to +\infty} X\big(u(t,u^{(0)}, 0), 0\big)=X(u^*,0)=0.$   
If this equilibrium $u_\infty$ is enough close to the starting pattern $\uzer$, then we classify this situation as \emph{recognition}. 
On the other hand, if the equilibrium $u_\infty$ is far from the starting pattern $\uzer$, we talk about \emph{free association}. 
%
%
%
\item  If the initial pattern is outside of a stability basin, the system may still exhibit oscillations due to the asymmetry of the interaction matrix, resulting in limit cycles rather than convergence (see Figure \ref{fig:asym_feat_1}). 
\item 
Otherwise the dynamics may alter the synaptic network creating a brand new equilibrium \((u_\infty, \xi_\infty)\) such that
\[
\lim_{t\to\infty} u(t,u^{(0)},\xi(\cdot)) = u_\infty, \qquad X(u_\infty, \xi_\infty) = 0. 
\]
We state that a new pattern $u_\infty$ 
has been \emph{recorded}, establishing a new equilibrium  altering the synaptic matrix $A$ with the addition of $\xi_\infty$. As already remarked in $(i)$, $u_\infty$ may be  close or far from $\uzer$. Then we will talk about recording or free associating, respectively.
In both cases, the system has learned/recorded something new.
%
%

\item %
The model may struggle to reach existing equilibria or even to create new equilibria activating the controls, ie., 
\[
\lim_{t \to +\infty} X(u(t, u^{(0)}, \xi), \xi(t)) \neq 0.
\]
Then \(\lim_{t \to +\infty} u(t, u^{(0)},\xi(\cdot)) \) does not exist, and we say \(u^{(0)}\) is neither recognized nor recorded. This scenario is likely to be quite common, 
and we refer to this as aimless \emph{ wandering}.
\item 
Finally, the optimal control model encompasses also the case in which a first incoming pattern leads to a network alteration deleting an existing equilibrium. 
Indeed assume that a synaptic matrix $A$ has been iteratively altered because of the submission of a number of incoming patterns $u^{(0)}, \dots, u^{(n_1)}$, and at the moment the resulting matrix is 
\[
A_{n_1} = A + \sum_{\alpha=1}^{n_1} \xi_\infty^{(\alpha)}. 
\]
where \(\xi_\infty^{\alpha}\) is the asymptotic control obtained submitting the pattern $u^{(\alpha)}$ at the $\alpha$--th step.
Let $\bar u$ be an equilibrium of the dynamics $A_{n_1}$. 
If a further sequence of inputs $u^{(n_1+1)}, \dots, u^{(n_2)}$ modifies the synaptic matrix beyond the bounds \(K\), i.e., 
\[
	A_{n_2} = A + \sum_{\alpha=1}^{n_1} \xi_\infty^{(\alpha)} + \sum_{\beta=n_1+1}^{n_2} \xi_\infty^{(\beta)}, \qquad 
\abs{\sum_{\beta=n_1+1}^{n_2} \xi_\infty^{(\beta)}} > K,
\]
then the previously recorded \(\bar{u}\) may no longer be reachable within a single step:  the pattern \(\bar{u}\) has been \emph{forgotten}. 
Of course further contributions $u^{(n_2+1)}, \dots, u^{(n_3)}$  may allow the recovering of \(\bar{u}\) as an equilibrium. We refer to this other situation as \emph{forgetting and restoring}.
\end{enumerate}

\begin{figure}[htbp] \begin{center} \includegraphics[width=60mm]{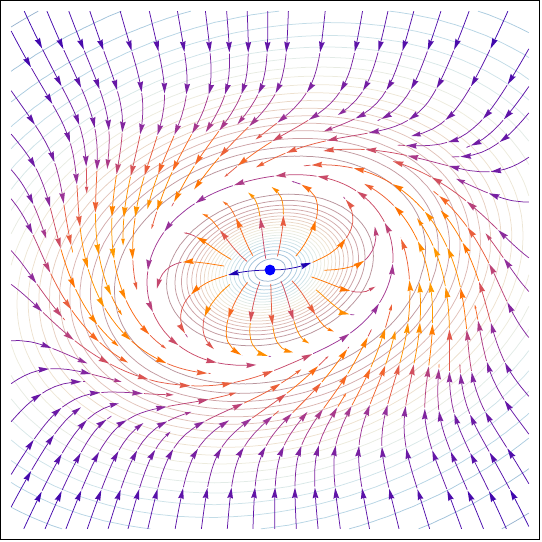} \caption{Example trajectories in the optimally controlled neural network model. Asymmetry may lead to limit cycles, resulting in oscillations.} \label{fig:asym_feat_1} 
\end{center} 
\end{figure}
From an operational standpoint, we can empirically estimate the upper limit $K$ for the synaptic updates $\xi$: a larger $K$ facilitates the construction and attainment of equilibria, while a progressively smaller $K$ inhibits this capability. 
\end{rem}

\begin{rem} The integral functional $J$ in (\ref{eq:varprin}), through a change of the time integration variable,
\[ [0,+\infty)\ni t \longmapsto s(t)=\frac{1-e^{-\lambda t}}{\lambda}\in \left[0, 1/\lambda \right),\]
is linked to the \emph{mean value} of $\ell$: 
%
\begin{eqnarray*}
	J(u^{(0)}, \xi) & = \int_{0}^{+\infty} \left(|X(u(t, u^{(0)}, \xi), \xi(t))|^2 + 
	|\xi(t)|^2\right)e^{-\lambda t} dt  \\ &= \int_{0}^{+\infty} \ell(t) e^{-\lambda t} dt 
	 = \int_{0}^{\frac{1}{\lambda}} \ell(t(s)) ds. 
\end{eqnarray*} 
Thus,
$$ 
\lambda J=\lambda\int_{s=0} ^{s=\frac{1}{\lambda}}\ell(t(s)) \, ds=\langle \ell(t(\cdot))    \rangle_{[0,1/\lambda]}.
$$
As $\lambda \to 0$, $s(t)$ approaches $s=t$ and $\lambda J$ converges to the mean value of $\ell$ over $[0,+\infty)$, reflecting a variation of the Final Value Theorem. 
\end{rem}


\subsection{Hamilton-Jacobi-Bellman Equation}
The following discussion primarily follows Bardi and Capuzzo-Dolcetta \cite{BCD}. We focus on the problem defined by the differential constraint (\ref{dynamics}):
\begin{equation}\label{eq:x_vs_u0}
	\dot u_i=(A_{ij}+\xi_{ij})g(u_j)-u_i,\quad u(0)=x,\ \ |\xi(t)|\leq (K,k),
\end{equation}
where $x$ represents the initial condition $u(0)$.

Our goal is to determine the control $\xi(t)$ that minimizes the cost functional $J$: 
\begin{equation} \label{eq:costfun}\fl
 v(x):=\inf_{|\xi(\cdot)|\leq (K,k)} J(x,\xi) 
 = \inf_{|\xi(\cdot)|\leq (K,k)}\int_0^{+\infty} \underbrace{\left( |X\big(u(t,x, \xi), \xi(t)\big)|^2+|\xi(t)|^2\right)}_{\ell(t)} e^{-\lambda t}\,dt\, ,
\end{equation}
where $ v(x)$ is usually named \textit{value function}.

If  $x$ is an equilibrium for $\xi=0$ (i.e., $X(x, 0) = 0$), it follows that $v(x) = 0$. Under the conditions 
(2.9-10), p.\,104 and $A_0,\dots,A_4$ 
specified in \cite{BCD}, we establish that $v(x)$  is a Lipschitz-continuous viscosity solution of the \emph{Hamilton-Jacobi-Bellman} (HJB) equation:
\begin{equation}\label{HJB}
    \lambda v(x)+H(x, \nabla v(x))=0,
\end{equation}
where the Hamiltonian function is:
\begin{equation}\label{Ham}
\eqalign{ H(x, p) &:=\sup_{|\xi|\leq (K,k)} \set{-X(x,\xi)\cdot p - \ell(X(x,\xi), \xi)} \\ 
& = -\inf_{|\xi|\leq (K,k)}\big(X(x,\xi)\cdot p+\ell(X(x,\xi), \xi)\big).}
\end{equation}

\subsection{Dynamic Programming Principle}
The Dynamic Programming Principle, as outlined in Bardi and Capuzzo-Dolcetta \cite[Relation (2.5)]{BCD}, states that for the optimal control $\xi^*$:
\begin{equation}\label{v}
    v(x)=\int_0^{t}\ell(u(s,x,\xi^*), \xi^*(s))e^{-\lambda s}\, ds+v(u(t,x,\xi^*))e^{-\lambda t}.
\end{equation}
This relation holds for any time $t\geq 0$.
By manipulating this equation, we obtain:
\begin{equation*}   \fl 
 v(u(t,x,\xi^*))e^{-\lambda t}\Big|_{t=0}-v(u(t,x,\xi^*))e^{-\lambda t}\Big|_{t>0}  
 =\int_0^{t}\ell(u(s,x ,\xi^*), \xi^*(s))e^{-\lambda s} ds\,,
\end{equation*}
Thus, we can express the integral as:
$$
\int_0^{t}\left(
\frac{d}{dt}\left(
v(u(s,x, \xi))e^{-\lambda s}\right)
+
\ell(u(s,x, \xi), \xi(s))e^{-\lambda s}\right)\, ds
=0\,
$$
for almost every $t\geq 0$. This leads to the differential equation:
\begin{equation}\label{paralyap}
    \frac{d}{dt}\left(
v(u(t,x, \xi))e^{-\lambda t}\right)
=-
\ell(u(t,x, \xi), \xi(t))e^{-\lambda t}.
\end{equation}
The presence of a global Lyapunov function, as already observed, appears inconsistent with the actual behavior of the mind. Nevertheless, relation (\ref{paralyap}) indicates a weak Lyapunov phenomenon, where the $e^{-\lambda t}$-discounted value function $v$ decreases with a time-rate according with the negative discounted Lagrangian $-e^{-\lambda t} \ell$. Rewriting gives:
\begin{equation}\label{paralyap-1}
    \frac{d}{dt}v(u(t,x,\xi^*))=\lambda v(u(t,x,\xi^*))- \ell(u(t,x,\xi^*), \xi^*(t))\,.
\end{equation}

Moreover, under the assumption that as $t\to +\infty$ we approach an equilibrium $x^*$,  (with $X(x^*, 0)=0$ and $\xi^*(t)$ not identically zero):
$$
\lim_{t\to+\infty} u(t,x,\xi^*)=x^*\,, \ \lim_{t\to+\infty}\xi^*(t)=0\,, \ X(x^*, 0)=0\,,
$$
the expected asymptotic behavior of the time-rate of $v$ is vanishing:
\begin{equation}   \fl 
    \lim_{t\to+\infty}\frac{d}{dt}v(u(t,x,\xi^*)) 
= \lim_{t\to+\infty}\left(\lambda v(u(t,x,\xi^*))- \ell(u(t,x,\xi^*)\,, \xi^*(t))\right)=0\,,
\end{equation}
since $\ell(x^*,0)=0$ and $v(x^*)=0$.

Finally, we can interpret the relation (\ref{paralyap}) using the so-called Witten deformation derivative \cite{AK,W}, as discussed in the literature. We define:
$$
D_t^{(\lambda)} f:=\Big(e^{\lambda t}  \frac{d}{dt} e^{-\lambda t}\Big) f,
$$
which allows us to rewrite  (\ref{paralyap}) as:
\begin{equation}
D_t^{(\lambda)} v=-\ell\leq 0\,.
\end{equation}
The Witten derivative  intertwines the value function with the convergence parameter $\lambda$;  
in fact, if we were to set $\lambda=0$ in the functional (\ref{eq:costfun}), 
the value function would effectively become a Lyapunov function. 
However, this is unrealistic, because it leads to a divergent functional. 
The fact that $\lambda$ cannot be zero, theoretically measures  
how the value function deviates from being a Lyapunov function.

\section{Pareto Optimization of a Multiobjective Cost Functional}

In the cost functional to be minimized (\ref{eq:varprin}):  
\begin{equation} \fl 
\inf_{|\xi(\cdot)|\leq (K,k)}
J(x,\xi) =\inf_{|\xi(\cdot)|\leq (K,k)}\int_0^{+\infty} \underbrace{\left( |X\big(u(t,x, \xi), \xi(t)\big)|^2+|\xi(t)|^2\right)}_{\ell(t)} e^{-\lambda t}\,dt,
\end{equation}
we can identify two competing objectives: 
\begin{enumerate} 
	\item Minimizing the norm of the vector field $X$ to quickly approach equilibria. 
	\item Minimizing the activation of the control field $|\xi|$ during the search for equilibria. \end{enumerate}
The trade-off between these objectives can be controlled by introducing weighting parameters $\alpha,\beta \in [0,1]$, $\alpha+\beta=1$, and reformulating the cost functional as follows: 
\begin{eqnarray}\label{pareto} 
	\eqalign{J_{\alpha,\beta} (x,\xi) = 
	\int_0^{+\infty} 
	\quadre{ \ell_{\alpha,\beta} (t)} e^{-\lambda t} dt, \ \\
	\mathrm{where, } \quad \hfill \ 
	\ell_{\alpha,\beta} (t) := \alpha |X\big(u(t,x, \xi), \xi(t)\big)|^2+ \beta |\xi(t)|^2. }
\end{eqnarray}
This suggests that the brain's wide range of possible reactions and outcomes in different situations, which can vary moment to moment, may result from an optimization process involving multiple objectives. The diversity in outcomes could thus reflect a set of Pareto optima.
When $\alpha$ is close to $0$ (and $\beta$ is close to $1$), the cost penalizes controls $\xi$ more, favoring approaches to existing equilibria (i.e., recognizing existing patterns). Conversely, larger values of $\alpha$ (and smaller of $\beta$) allow for larger controls $\xi$, promoting the exploration of new equilibria that do not exist in the original configuration of $X$.
%
%
A coupling between optimal control theory and Pareto multiobjective optimization has been explored in \cite{LC}.

\section{Synopsis of the standard Hopfield model}\label{sec:resume}
We lay down a brief survey on the classical Hopfield model, in the original discrete version \cite{H} and in its continuous version \cite{H1}.
\begin{description}
\item[$N$:] number of neurons,
\item[$T_{ij}$:] synaptic interaction symmetric matrix, $T_{ii}=0,\ \ i,j=1,\dots, N$,
\item[$V_i$:] the status of the $i$-th neuron, i.e., $V_i=1$: firing,\ $V_i=0$: not firing,
\item[$u_i$:] the electric potential in the $i$-th neuron, i.e., $u_i^{(1)} = \sum_{j=1}^N T_{ij}V_j ^{(0)}$,
\item[$g_0$:]  when on the $i$-th neuron it arrives a resulting potential $u_i\in{\mathbb R}$,
then its status will be $V_i=g_0(u_i)$, where, for a (small) threshold $a>0$ $V_i=g_0(u_i)=1$ if $u_i\geq a$,
and $V_i=g_0(u_i)=0$ if $u_i<a$. 
Given an initial pattern $V_i^{(0)}$, by the matrix $T_{ij}$ the brain interacts with itself, giving the new status $V^{(1)}_i=g_{0}\tonde{u_i^{(1)}}=g_0\tonde{\sum_{j=1}^N T_{ij}V_j ^{(0)}}$, this mechanism is iterated, arriving to define the
\item[\textbf{ Discrete Dynamics} \cite{H}:] \hspace{20mm} \hfill $\displaystyle V^{(n+1)}_i= g_0\tonde{ u_i^{(n+1)}}=g_0\tonde{\sum_{j=1}^N T_{ij}V_j^{(n)}}.$ \\ We can also write: 
\[\displaystyle V^{(n+1)}_i-V_j^{(n)} = g_0\tonde{\sum_{j=1}^N T_{ij}V_j^{(n)}}-V_j^{(n)}, \qquad \mathrm{or,}\]
\[ u_i^{(n+1)}- u_i^{(n)} = \sum_{j=1}^N T_{ij}g_0(u_j^{(n)})- u_i^{(n)} .\]
In literature, the second equation is usually interpreted as the time one step of the finite reduction of an \textsc{ode}. To obtain such an \textsc{ode}, we introduce the `sigmoid' function, see  Fig.\,\ref{fig:sigmoid}:
\[g_\varepsilon : {\mathbb R}\to[0,1], \quad \mathrm{e.g.,} \quad g_\varepsilon (x)=\frac{1}{\pi}\arctan \tonde{\frac{x}{\varepsilon}}+\frac{1}{2},\]
where $\varepsilon>0$ is a fixed small parameter.
\begin{figure}[htbp]
\begin{center}
\begin{tikzpicture}
\begin{axis}[
    domain=-15:15,
    xscale=1.,yscale=0.5,
    ylabel= $g_{\varepsilon}(x)$,
    xlabel= $x$,
    xmin=-15, xmax=15,
    ymin=-0.2, ymax= 1.5,
    xtick={0.01},
    xticklabels={     
      	0
    },
    ytick={ 1.0},
    yticklabels={     
    	1 
    },
    x tick label style={xshift={(\tick==0.01)*.6em}},
    samples=1000,
    axis lines=center
]
    \addplot+[mark=none,color=red,thick] {(1/pi)*rad(atan(5*x))+0.5};
    \addplot+[mark=none,color=blue,dashed,domain=0:10] {1};
\end{axis}
\end{tikzpicture} 
\caption{Sigmoid.
}
\label{fig:sigmoid}
\end{center}
\end{figure}
As a result we obtain:
\item[\textbf{Continuous Dynamics \cite{H1}}:] \hspace{5mm} \hfill $\displaystyle \frac{du_i }{dt} =\sum_{j=1}^N T_{ij}g_\varepsilon(u_j)-u_i,$
where we mean $V_i=g_\varepsilon(u_i)$ and $u_i=g_{\varepsilon}^{-1}(V_i)$. 
\item[\textbf{Energy}:] \hspace{5mm} $\displaystyle E(V):= -\frac{1}{2}\sum_{i,j=1}^N T_{ij}V_iV_j+\sum_{i=1}\int_0^{V_i}g_{\varepsilon}^{-1}(x)dx$ 

\begin{equation}\eqalign{
 \dot E & = - \sum_{i=1}^N \tonde{\sum_{j=1}T_{ij}V_j- u_i}\dot V_i=-\sum_{i=1}^N\dot u_i\dot V_i  \\
 &= -\sum_{i=1}^N\underbrace{ \frac{d}{dx}     g_{\varepsilon}(x)|_{x=u_i} }_{>0} X_i(u)^2 \leq 0\,.}
\end{equation}

Note that
\begin{enumerate}[(i)]
\item $E$  is lower bounded,
\item The definition of $E$ shows that Lyapunov behavior occurs only for symmetric $T$.
\end{enumerate}
\end{description}


\section{Numerical simulation of the optimally controlled Hopfield network}
\label{sec:numerics}

We consider a starting model composed by two neurons, which is enough simple for a thorough investigation and at the same time is able to reproduce the interesting behaviour phenomena. 
The instantaneous electric potential pattern measured in each neuron is therefore a two dimensional vector $u=(u^1,u^2)\in\R^2$. 

\subsection{Time one discretization}

We set $n$ as the number of time intervals in which the time window observed is divided. We denote by $t_{0},t_{1},\dots,t_{n}$ the time instants and by $u_{k} = u(t_{k})$ the electric potential pattern at time $t_{k}$. $u_{0}$ is the incoming pattern.  
If we adopt a time one discretization we have $t_k = k$, and the 
network discrete updating is given by 
\[ 
	u_{k+1} = (A + \xi_{k}) \cdot g_\epsilon(u_{k}),
\]
where $\xi_k:=\xi(t_k)$ is the control applied to the system in the $k$-th time instant. 
The vector field $X(u_k)$ at time $t_k$ is 
\begin{equation}
	X(u_k) = u_{k+1}-u_{k} = (A + \xi_k) \cdot g_\epsilon(u_{k}) - u_{k}.
\end{equation}
%
Now we can write the discrete version of the cost functional to be minimized as 
\begin{equation}\label{eq:jfunc}
\eqalign{\fl J(u,\xi) :=   \exp(-\lambda 0)  \alpha  \norm{u_{1}-u_{0}}^2 +\\ 
	+\sum_{k=0,\dots,n-1}             \exp(-\lambda k)  \alpha  \norm{u_{k+1}-u_{k}}^2 +\\  + 
             \sum_{k=1,\dots,n}\exp(-\lambda k) \beta \norm{\xi_k}^2,  \qquad \alpha, \beta \in \R^{>0}.}
\end{equation}
 

We consider as zero-model the dynamics obtained with the fixed matrix $A$.  In this case, every starting point $u_0$ is led to an existing equilibrium of the original dynamical system of the zero-model. 

\subsection{Symmetric positive matrix. Tradeoff dynamics with $0\leq \alpha\leq 1.$ and $\beta=1-\alpha$}

We want to study how the dynamics changes as the values of the coefficients $\alpha$ and $\beta$ in the functional (\ref{eq:jfunc}) 
change, in particular we want to start with $\beta=1$ and $\alpha=0$, which produce the static matrix dynamics $\xi\equiv 0$, and observe what happens as $\beta$ decreases and $\alpha$ increases at the same time, while $\alpha+\beta=1$. 
When $\beta\gg 0$, we expect that the equilibrium is not altered, however, as $\beta$  decreases towards $0$, the optimal path to the equilibrium could be modified, probably shortening, in general. 
Finally, when $\beta$ is close to zero, we expect that the equilibrium itself modifies, most likely getting closer to the starting point $u_0$ in general. 
In this first round of experiments we neglect the contribution of the large time discount $e^{-\lambda t}$ setting $\lambda = 0$, 
because in all the experiments conducted convergence seems reached within the time interval of 50 time steps considered. 



\begin{figure}[htbp]
\begin{center}
\begin{tabular}{|c|c|c|}
\hline
\includegraphics[width=45mm]{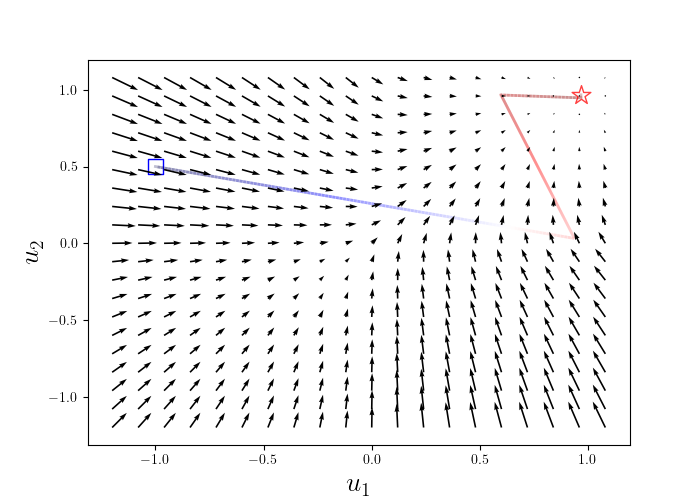} &
\includegraphics[width=45mm]{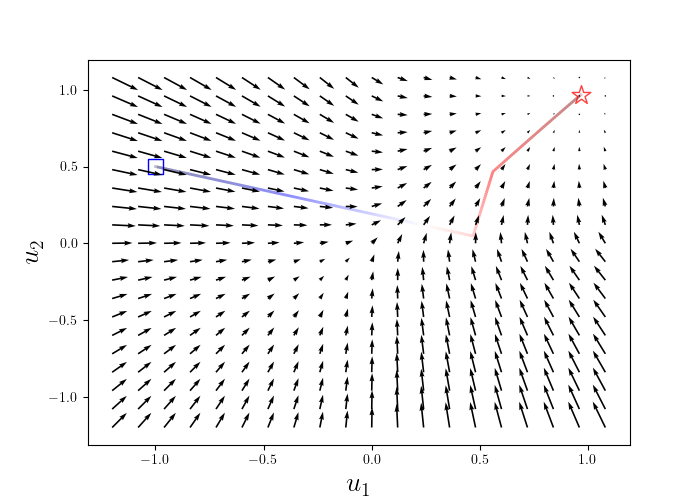} &
\includegraphics[width=45mm]{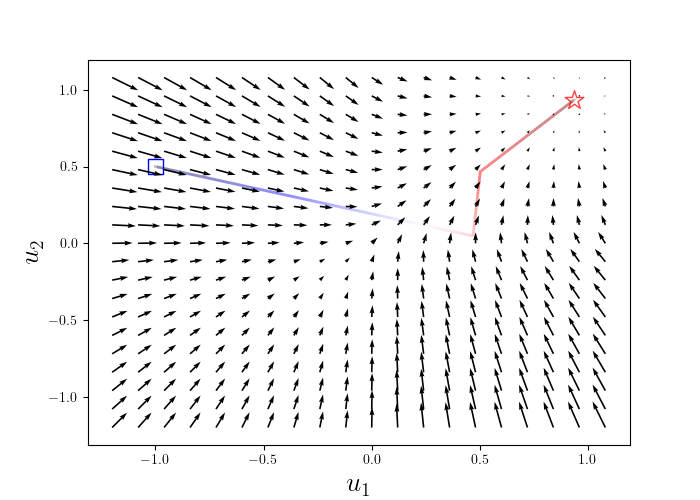} \\\hline
\includegraphics[width=45mm]{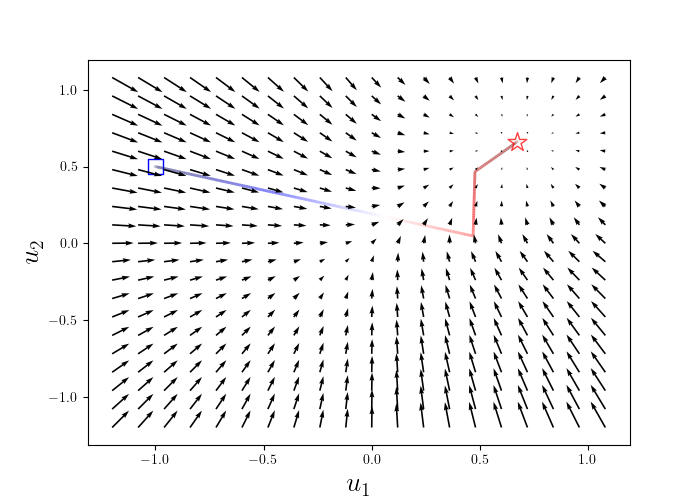} &
\includegraphics[width=45mm]{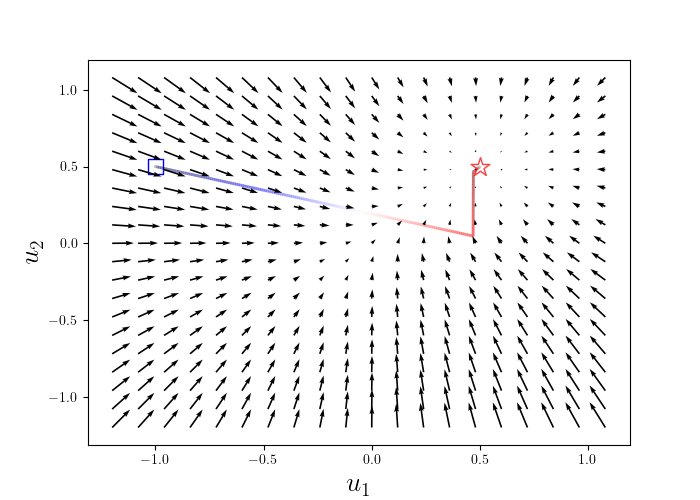} &
\includegraphics[width=45mm]{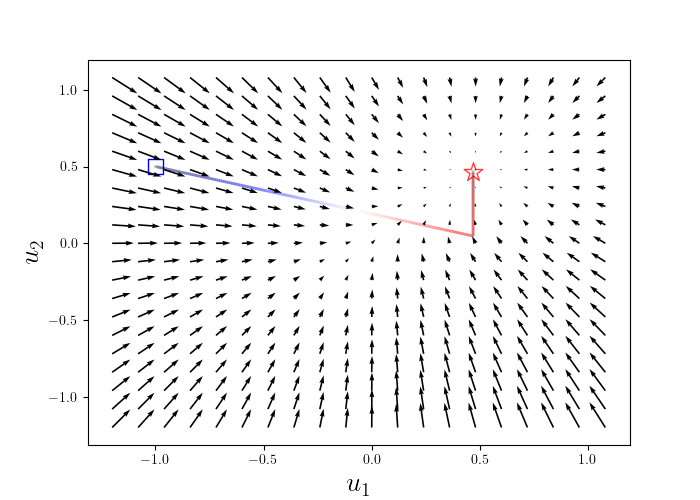} \\\hline
\end{tabular}
\includegraphics[width=125mm]{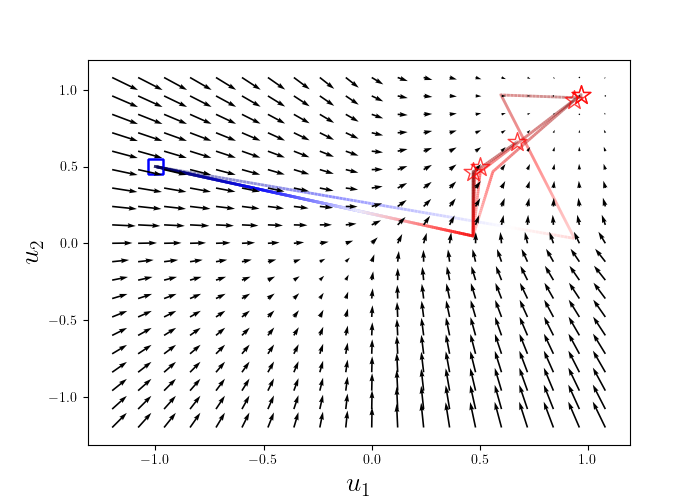} 
\caption{Optimally controlled dynamics as $\beta$ varies from $1$ to $0$ and $\alpha=1-\beta$. The blue square is the starting point while the red star is the equilibrium, i.e., the limit of the dynamics for $t\to\infty$. The color of trajectory ranges from blue to red as the time count increases. In the lower panel all the trajectories are plotted at the same time.}
\label{figN1:betatradeoff}
\end{center}
\end{figure}

We consider for simplicity a starting positive symmetric matrix $A= \tonde{
\begin{array}{cc}
 0 & 1 \\ 1 & 0
\end{array}}$ and define $g_\epsilon$ with $\epsilon= 0.1$. The matrix of controls is simular to $A$ in the sense that we take zero elements in the main diagonal matrix $\xi := \tonde{\begin{array}{cc}  0 & \xi^1 \\ \xi^2 & 0 \end{array}}$. 
The starting point is $u_0=(-1,0.5)$. As expected we see that for values of $\beta$ sensibly larger than zero the equilibrium of the zero model is also the equilibrium of the optimally controlled dynamics. As $\beta$ approaches to zero firstly the path is shortened and finally the equilibrium $u_\infty$ itself moves closer to the starting pattern $u_0$ (see Figure \ref{figN1:betatradeoff}).

\subsection{Symmetric positive matrix. Starting from different points}

\begin{figure}[htbp]
\begin{center}
 \includegraphics[width=125mm]{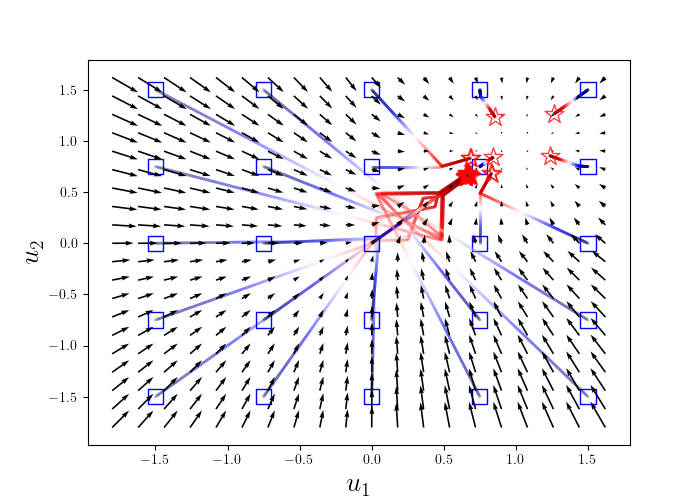} 
\caption{Optimally controlled dynamics as $\beta=0.0005$, $\alpha=0.9995$ and the starting pattern $u_0$ varies in a regular grid.}
\label{figN2:multistart}
\end{center}
\end{figure}

We fix the values of $\beta$ and $\alpha$ to $0.005$ and $0.995$ respectively and observe how the equilibria are altered as the starting point is taken from the regular grid of five levels and for bounds equal to $-1.5$ and $1.5$. The same behavior observed above is realized, i.e., the equilibrium is closer to the starting point (see Figure \ref{figN2:multistart}).

\subsection{Symmetric positive matrix. Incremental learning process. }

At the end of the previous process, for $\beta=0.005$, a new equilibrium is determined cancelling the previously existing one. We fix then the starting matrix as $A+\xi_\infty$ and set the corresponding dynamics as the new zero model. 

We observe what happens when we iterate the same process, always with the same starting potential $u_0$  updating the fixed matrix with the optimal controls, consolidating the learning process at each step. 

At each iteration the newly generated equilibrium gets closer to the starting point $u_0$.

\begin{figure}[htbp]
\begin{center}
 \includegraphics[width=125mm]{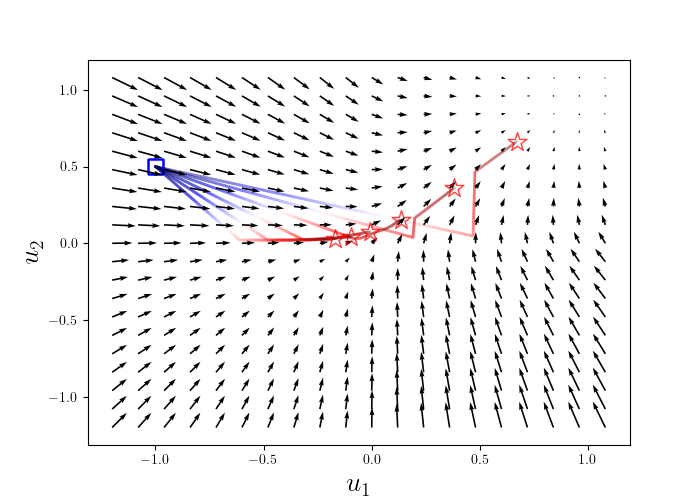} 
\caption{Optimally controlled dynamics iterated 6 times starting every time from the same $u_0$. At every iteration the same starting electric pattern $u_0$ is proposed to the network. The matrix connectivity $T$ is updated at every step by adding to the matrix $A_n$ of the previous step the control matrix $\xi_\infty$ obtained in the limit $t\to\infty$. Every new updated matrix $A_{n+1}$ has a dynamics converging to a new equilibrium closer to the starting pattern $u_0$. }
\label{figN3:restart}
\end{center}
\end{figure}

\subsection{Symmetric positive matrix. Aimless wandering dynamics as $\lambda>0$. }

The large time discount term $e^{-\lambda t}$ ensures theoretical convergence of (\ref{eq:jfunc}) in the search of the optimal controls, this implies that the deviation from zero for both $\abs{X(k)}=\abs{u_{k+1}-u_k}$ and $\abs{\xi_k}$ itself are less and less penalized as time passes. Such insensitivity in the controls for large times allows for the dynamics to become less and less predictable and sensitive to perturbations. 

\begin{figure}[htbp]
\begin{center}
 \includegraphics[width=125mm]{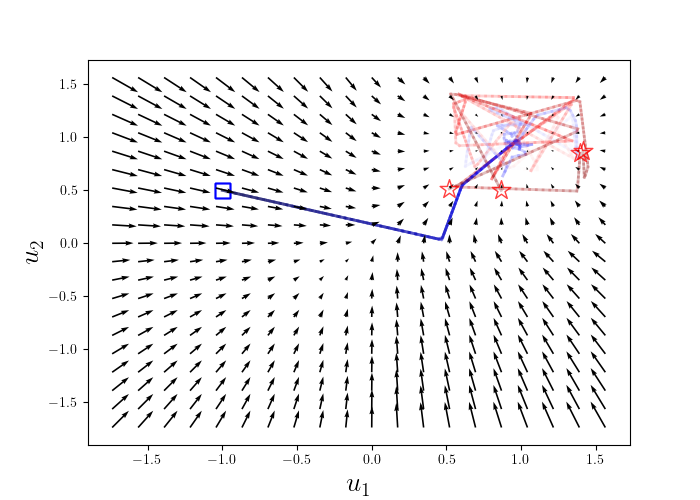} 
\caption{Effects of the introduction of discount term $e^{-\lambda t}$. The functional (\ref{eq:jfunc}) becomes insensitive to the later values of both $X-\xi$ and $\xi$ causing an unpredictable behaviour of the trajectory for large times. Re-running several trajectories starting from the same pattern $u_0$ produce trajectories identical for the first steps but later on the trajectories diverge one from the other occupying a growing size cloud around the equilibrium of the zero model. We call this behavior ``wandering''.}
\label{figN4:wandering}
\end{center}
\end{figure}

We consider what happens to the dynamics as $\lambda $ increases from $0.0$ to $0.5$ with fixed values of $\alpha$ and $\beta$. 
As it can be observed, the final points occupy a growing size cloud of points approximately centred  on the equilibrium obtained with $\lambda=0.$ (see Figure \ref{figN4:wandering}).

The code used for the numerical experiments is available at: \\
\texttt{https://github.com/aruberuto-rouison/OCHNeuralNetworks}

\section*{Acknowledgement} The authors warmly thank Martino Bardi for kind discussions and suggestions.
The contribution of F.\ Cardin to this paper has been realised within the sphere of activities of the GNFM of INDAM. The Italian ministry provided a financial support for A. Lovison through grants MIUR PRIN 2017KL4EF3 and 2020F3NCPX.

\end{document}